# The Symmetry Properties of a Non-Linear Relativistic Wave Equation: Lorentz Covariance, Gauge Invariance and Poincare Transformation

A. M. Awobode
Department of Physics, University of Illinois, Urbana-Champaign, IL 61801, U.S.A

**Abstract**

The Lorentz covariance of a non-linear, time-dependent relativistic wave equation is demonstrated; the equation has recently been shown to have highly interesting and significant empirical consequences. It is established here that an operator already exists which ensures the relativistic properties of the equation. Furthermore, we show that the time-dependent equation is gauge invariant. The equation however, breaks Poincare symmetry via time translation in a way consistent with its physical interpretation. It is also shown herein that the Casimir invariant $P_\mu P^\mu$ of the Poincare group, which corresponds to the square of the rest mass $M^2$ can be expressed in terms of quaternions such that M is described by an operator Q which has a constant norm $\|Q\|$ and a phase $\vartheta$ which varies in hypercomplex space.

## 1. Introduction: Relativistic Invariance of the Wave Equations

The principle of special relativity requires that the form of the laws of physics remain invariant for all observers in relative uniform motion. In both classical and quantum theories, the impact of the relativity principle has been quite prodigious and the conformity with the relativity principle of a physical law, can be clearly demonstrated by its invariance under the homogenous Lorentz group; a Lorentz transformation can equivalently be regarded as a rotation through an imaginary angle in a four dimensional space-time continuum. An extension of the Dirac equation, which corrects some well-known deficiencies of the theory and from which the electron/muon anomalous gyro-magnetic factors have recently been calculated [7], was earlier proposed [1]; the calculated values of the orbital and spin g-factors are, to four significant figures, in good agreement with experimental data thus creating renewed interest in the modified theory. The proposed equation contains an additional time-dependent mass term obtained by considerations based on the respected facts of relativity and quantum theory [2]. However, since the Dirac equation is demonstrably relativistic, it is necessary to show that the additional term does not break or destroy its invariance under Lorentz transformations.

The relativistic invariance of a free-particle Dirac equation

$$\left(i\hbar\gamma_\mu \frac{\partial}{\partial x_\mu} + mc^2\right)\psi = 0 \qquad (1.1)$$

is established for example, by showing that an operator S exists satisfying the equation,
$$S^{-1}\gamma_\mu S = a_{\mu\nu}\gamma_\mu \quad (1.2)$$
where $a_{\mu\nu}$ is the 4 x 4 Lorentz transformation matrix and $\gamma_\mu$ are the Dirac matrices [3,4].

In general, the appropriate S is given as

$$S = \exp(1/2\,\omega\,\bar{n}\alpha) \quad (1.3)$$

where $\bar{n}$ is the unit vector in the direction of motion, and $\alpha_k$ is the anti-Hermitian Dirac matrices $\gamma_k\gamma_0$ [4]. For a frame of reference moving with velocity v = βc along the x-axis, the operator S reduces to the familiar

$$S_{Lor} = \exp(-1/2\,\omega\sigma_{01}) = \cosh(\omega/2) - i\gamma_3\gamma_4 \sinh(\omega/2) \quad (1.4)$$

where tanh ω = v/c.

When the time-dependence of the rest mass is taken into considerations, the relativistic wave equation for spin-1/2 particles in the absence of external fields becomes

$$\left(i\hbar\gamma_\mu \frac{\partial}{\partial x_\mu} + mc^2\right)\psi = mc^2 \exp(2i\alpha \cdot pct/\hbar)\psi \quad (1.5)$$

This equation has been found useful in the resolution of some formal and empirical difficulties of the Dirac theory [5]. It is thus necessary to explicitly demonstrate the covariance of the equation under a Lorentz transformation.

We shall proceed as was done for the Dirac equation. In the primed frame of reference, the equation (1.5) is transformed into

$$\left(i\hbar\gamma_\mu \frac{\partial}{\partial x'_\mu} + mc^2\right)\psi'(x') = mc^2 \exp(2i\alpha \cdot p'ct'/\hbar)\psi'(x') \quad (1.6)$$

where the physical constants h, c and m are invariant under Lorentz transformation, and also the matrices $\gamma_\mu$ are unaffected by relative motion.

Because $x'_\mu = a_{\mu\nu}x_\nu$ by Lorentz transformation, eqn (2.6) above becomes

$$i\hbar\gamma_\mu a_{\mu\nu} \frac{\partial \psi'(x')}{\partial x'_\mu} + mc^2\psi'(x') = mc^2 \exp(2i\alpha \cdot p'ct'/\hbar)\psi'(x') \quad (1.7)$$

The term $\exp(2i\alpha \cdot p'ct'/\hbar)$ is formally analogous to an undulating phenomenon, the phase of which, according to relativistic kinematics is invariant [6]. i.e

$$\exp(2i\alpha \cdot p'ct'/\hbar) = \exp(2i\alpha \cdot pct/\hbar) \quad (1.8)$$

We require an S such that $S\psi(x) = \psi'(x')$, and so we obtain

$$i\hbar \gamma_\mu a_{\mu\nu} S \frac{\partial \psi}{\partial x_\mu} + mc^2 S\psi = mc^2 \exp(2i\alpha \cdot pct/\hbar) S\psi \qquad (1.9)$$

Pre-multiplying through by $S^{-1}$ we have

$$i\hbar S^{-1} \gamma_\mu a_{\mu\nu} S \frac{\partial \psi}{\partial x_\nu} + mc^2 \psi = mc^2 S^{-1} \exp(2i\alpha \cdot pct/\hbar) S\psi \qquad (1.10)$$

As before (1.9) is covariant if S exists such that the following two conditions

$$S^{-1} \gamma_\mu a_{\mu\nu} S = \gamma_\nu \qquad (1.11)$$

and $S^{-1} \exp(2i\alpha \cdot pct/\hbar) S = \exp(2i\alpha \cdot pct)$, i.e.,

$$[\exp(2i\alpha \cdot pct/\hbar), S] = 0 \qquad (1.12)$$

are satisfied. We require that S commutes with $\exp(2i\alpha \cdot pct/\hbar)$ in order to simplify the right hand side of eqn. (1.9)

Thus if we can find an S which simultaneously satisfies eqns (1.10) and (1.11) we would have proven that the postulated wave equation (1.5) is Lorentz covariant. It will be shown that the matrix S given in eqn (1.3) satisfies both conditions. Eqn (1.10) is the same as the condition (1.2) for the covariance of eqn (1.1); thus it suffices to show that the S for the free particle Dirac equation, commutes with the added operator as stated in eqn (1.11), in order to prove the Lorentz covariance of (1.5).

The matrix S in eqn (1.3) can be written as

$$S = \exp(1/2\omega \cdot p/|p|) \qquad (1.13)$$

where the unit vector n is in the direction of motion p, i.e. n = p/|p|. We observe therefore that since

$$[\exp(2i\alpha \cdot pct), \exp(i\omega\alpha \cdot p/2|p|)] = 0 \qquad (1.14)$$

matrix addition is commutative and $[\alpha\cdot p, \alpha\cdot p] = 0$, thus establishing eqn (1.11) and hence proving that (1.5) is invariant under Lorentz transformations. This is perhaps more clearly demonstrated by putting $A \equiv 2ict/\hbar(\alpha\cdot p)$ and $B \equiv i\omega/2|p|(\alpha\cdot p)$ such that

$$[e^A, e^B] = e^A e^B - e^B e^A = e^C - e^{C'}$$

We note that $e^A e^B = e^{A+B+(1/2)[A,B]}$ for any operators A and B. However if $[A, B] = 0$, then $e^C = e^{C'}$ and therefore

$$[e^A, e^B] = 0$$

which demonstrates (1.13), thus establishing (1.11) and hence proving that (1.5) is invariant under Lorentz transformations.

## 2. Gauge Invariance

The Dirac equation, when minimally coupled to the electromagnetic field as follows:

$$\left[ i\hbar\gamma_\mu \left( \frac{\partial}{\partial x_\mu} - \frac{e}{c} A_\mu \right) + mc^2 \right] \psi = 0 \tag{2.1}$$

is known to be invariant under the gauge transformation $A_\mu \to A'_\mu$, $\psi \to \psi'$ [3], where

$$A'_\mu = A_\mu + \frac{e}{c} \frac{\partial G}{\partial x_\mu} \tag{2.2a}$$

and

$$\psi' = \exp(ie/\hbar c) G \psi \tag{2.2b}$$

(In the above, G is a scalar function satisfying the zero mass Klein-Gordon equation and Maxwell's equations.) The invariance of the Dirac equation means that equation (2.1) is, apart from a phase factor, identical to,

$$\left[ i\hbar\gamma_\mu \left( \frac{\partial}{\partial x_\mu} - \frac{e}{c} A'_\mu \right) + mc^2 \right] \psi' = 0 \tag{2.3}$$

The extended Dirac equation with minimal coupling is

$$\left[ i\hbar\gamma_\mu \left( \frac{\partial}{\partial x_\mu} - \frac{e}{c} A_\mu \right) + mc^2 \right] \psi = mc^2 \exp(2i\alpha \cdot (p - eA/c) ct / \hbar) \psi \tag{2.4}$$

It is necessary to determine whether the additional operator which appears in the extended Dirac equation above preserves the gauge invariance of the theory or otherwise. That is, we should find out whether the expression

$$\left[ i\hbar\gamma_\mu \left( \frac{\partial}{\partial x_\mu} - \frac{e}{c} A'_\mu \right) + mc^2 \right] \psi' = mc^2 \exp(2i\alpha \cdot (p - eA'/c) ct / \hbar) \psi' \tag{2.5}$$

containing the transformed potentials $A'_\mu$ and the wavefunction $\psi'$ is (or is not) identical to (2.4).

First we consider the left-hand-side of eqn (2.5) and substitute $A_\mu$ and from (2.2) into it. We obtain

$$exp(ie/\hbar cG)\left[i\hbar\gamma_\mu\left(\frac{\partial}{\partial x_\mu}-\frac{e}{c}A_\mu\right)+mc^2\right]\psi \qquad (2.6)$$

Next we examine the right-hand-side of (2.5), i.e., we evaluate the action of the operator $exp[2i\alpha\cdot(p-eA'/c)ct/\hbar]$ on $\psi'$.

Let us put $(\alpha\cdot p-eA'/c)\equiv B$, $2ict/\hbar\equiv\lambda$ and note that

$$exp(\lambda B)=I+\lambda B+\frac{\lambda^2 B^2}{2!}+\frac{\lambda^3 B^3}{3!}+... \qquad (2.7)$$

where I is the identity operator. From (2.7) we observe that it is necessary to examine the action of the powers of B on $\psi'$. We shall therefore, in the first instance, evaluate

$$B\psi'=\alpha\cdot(p-eA'/c)\psi'=\alpha\cdot(p-eA/c-(e/c)\nabla G)exp(ieG/\hbar c)\psi \qquad (2.8)$$

Perfoming the differentiation $\mathbf{p}\psi=-i\hbar\nabla exp(ieG/\hbar c)=-(e/c)exp(ie/\hbar c)G\nabla G$, we see that

$$\alpha\cdot(p-eA'/c)\psi'=exp(ieG/\hbar c)\alpha\cdot(p-eA/c)\psi=exp(ieG/\hbar cQ)\psi \qquad (2.9)$$

where we have put $\alpha\cdot(p-eA/c)=Q$.

Likewise the action of the operator $B^2$ on $\psi'$ is obtained as follows:

$$\begin{aligned}B^2\psi'&=\alpha\cdot(p-eA'/c)^2\psi'=\alpha\cdot(p-eA/c-(e/c)\nabla G)\alpha\cdot(p-eA/c)\\&=exp(ie/\hbar c)[\alpha\cdot(p-eA/c)]^2\psi\end{aligned} \qquad (2.10)$$

i.e.,

$$B^2\psi'=exp(ieG/\hbar c)Q^2\psi \qquad (2.11)$$

Repeated application of B on $\psi'$ yields

$$B^n\psi'=exp(ie/\hbar c)[\alpha\cdot(p-eA/c)]^n\psi=exp(ie/\hbar c)Q^n\psi \qquad (2.12)$$

where n = 2, …, ∞

Therefore the action of $exp(\lambda B)$ on $\psi'$, using (2.7) and (2.12) is

$$\begin{aligned}exp(\lambda B)\psi'&=(I+\lambda B+\frac{\lambda^2 B^2}{2!}+\frac{\lambda^3 B^3}{3!}+...)\psi'\\&=exp[(ie/\hbar c)G]\sum_{k=0}^\infty\frac{\lambda^k}{k!}Q^k\psi=exp(ieG/\hbar c)exp(\lambda\alpha\cdot(p-eA/c)\psi\end{aligned} \qquad (2.13)$$

since

$$\sum_{k=0}^{\infty}\frac{\lambda^k}{k!}Q^k = \exp[2i\alpha\cdot(p-eA/c)]$$

We therefore find that the action of the operator $\exp[2i\alpha(p-eA'/c)ct/\hbar]$ on $\psi'$ is given as

$$\exp[2i\alpha\cdot(p-eA'/c)ct]\psi' = \exp(ieG/\hbar c)\exp[2i\alpha\cdot(p-eA/c)ct/\hbar]\psi \qquad (2.14)$$

Thus eqns (3.6) and (3.14) when put together gives

$$\exp(ieG/\hbar c)\left[i\hbar\gamma_\mu\left(\frac{\partial}{\partial x_\mu}-\frac{e}{c}A_\mu\right)+mc^2 - mc^2\exp[2i\alpha\cdot(p-eA/c)ct/\hbar]\right]\psi = 0$$

(2.15)

Since $\exp(ieG/\hbar c)$ is arbitrary, we may write

$$\left[i\hbar\gamma_\mu\left(\frac{\partial}{\partial x_\mu}-\frac{e}{c}A_\mu\right)+mc^2\right]\psi = mc^2\exp[2i\alpha\cdot(p-eA/c)ct/\hbar]\psi \qquad (2.16)$$

which is of the same form as eqn (2.4), thus establishing the gauge invariance of the proposed time-dependent relativistic wave equation.

## 3. Poincare Transformation

The time-dependent, modified Dirac equation (eqn 1.5) written as

$$i\hbar\frac{\partial\psi}{\partial t} = H\psi = [c\alpha\cdot p + \beta mc^2 - \beta mc^2\exp(2i\alpha\cdot pct/\hbar)]\psi \qquad (3.1)$$

is such that, under time translations $t \to t' = t+a$, $H \to H'$, but $H \neq H'$, i.e., the Hamiltonian is not invariant, because the term $\beta mc^2\exp(2i\alpha\cdot pct/\hbar)$, which couples the negative- and positive-energy states, does not conserve energy. This observation is consistent with the Heisenberg equation of motion; although H commutes with itself, $(i\hbar)dH/dt = [H,H] + \partial H/\partial t \neq 0$, since H has an explicit time dependence. As previously interpreted [3], the time-dependent rest-energy term (analogous to a time-dependent electromagnetic field) induces transitions between spinor states $u_{p,s} \leftrightarrow u_{p',s'}$, or alternatively, describes the creation & annihilation of electron-positron pairs within the Compton distance $x : \hbar/mc$, at an angular frequency $\omega : 2mc^2/\hbar$ [1].

As a consequence of the above the corrected Dirac equation which, as shown in section 1, is invariant under the homogeneous Lorentz transformation is <u>not</u> invariant under Poincare transformation; the latter appropriately describes the relativistic invariance of isolated systems, together with their symmetry with respect to space and time translations. The unitary irreducible representations of the Poincare group are equivalent to the wave equations which describe the properties of elementary particles in the absence of all interactions [3]. Nevertheless, for the description of interacting systems, the properties of the finite representations of the homogeneous Lorentz group are sufficient [8].

The two Casimir invariant operators of the homogeneous Lorentz group are

$$\mathbf{C}^2 - \mathbf{D}^2 = (1/2) C_{\mu\nu} C^{\mu\nu} \quad \text{and} \quad -\mathbf{C} \cdot \mathbf{D} = (1/8) \varepsilon^{\mu\nu\rho\sigma} C_{\mu\nu} C_{\rho\sigma} \tag{3.2}$$

where $\mathbf{C}$ and $\mathbf{D}$ are vectors formed from the generators $C_{\mu\nu}$ of the group. In an irreducible representation of the Lorentz group, the values of the quantities $\mathbf{C}$ and $\mathbf{D}$ are $-\{j(j+1) + j'(j'+1)\}I$ and $i\{j(j+1) - j'(j'+1)\}I$ respectively, where $I$ is the identity; the pair of indices $j, j'$ take the values 0, ½, 1, 3/2 …, and $j(j+1), j'(j'+1)$ are, respectively, the eigenvalues of the operators $\mathbf{J}^2, \mathbf{K}^2$, where $\mathbf{J} = i(\mathbf{C}+i\mathbf{D})/2$ and $\mathbf{K} = i(\mathbf{C}-i\mathbf{D})/2$. The Casimir invariant operators of the Poincare group, on the other hand, are

$$P_\mu P^\mu = M^2, \quad W_\mu W^\mu = S^2 \tag{3.3}$$

which correspond to the square of the mass and spin respectively; thus $M^2$ and $S^2$ are constant for isolated systems. That is, $M^2 = \Omega^2$ (a constant, real number), which implies that $M = \pm\Omega$ or $\Omega = \pm M$; it is equally valid to describe the mass as $\mathbf{M} = |\Omega| \exp(i\sigma_K \theta_L)$, i.e. $\mathbf{M}$ is a quantity in the hypercomplex planes (or space), having a variable phase $\theta_L$ and a constant norm $|\mathbf{M}|^2 = \Omega^2$ (see **Appendix**). Hence, the fact that the unmodified Dirac equation is invariant under Poincare transformations does not necessarily imply that the mass term is a fixed real number $|\Omega|$. The mass can, in general, be described by a quaternion Q with a constant absolute magnitude $\|Q\|$ and variable phase $\vartheta$. This description gives a codified classification of the possible irreducible representations of the Poincare group, labeled by the values of $\mathbf{M}$, which includes the real values $\Omega = \pm M$.

### 4. Conclusion

By establishing the existence of an operator S, it has been shown that the extended Dirac equation is manifestly Lorentz covariant. Also, the gauge invariance of the equation is demonstrated. Thus the principle of charge conservation is preserved in the modified Dirac theory and the form of the wave equation is the same for all inertial observers in relative motion. However, the extended Dirac equation is not invariant under Poincare transformation and, as consequence, displays some interesting properties.

The interaction term in eqn (3.1) breaks Poincare symmetry via time translation, but invariance under proper, restricted Lorentz transformation is preserved. Thus the equation, from a relativistic point of view, is satisfactory. Although the Dirac Hamiltonian $H_D = c\mathbf{\alpha}\cdot\mathbf{p} + \beta Mc^2$ is invariant under Poincare transformations, its mass term M can be consistently interpreted as an operator which can, in general, assume real, imaginary or complex values[1].

Symmetry principles, like all other laws of nature, are not *a-priori* self-evident. They have to be inferred from empirical data and then used to predict testable consequences which can be verified by observations. So far, no discrete or continuous space-time symmetry has been found which implies that the rest-energy $Mc^2$ is a conserved quantity, or that it is invariant with respect to coordinate time (i.e. a constant of the motion); such a group-theoretic result will be in direct contradiction with the notion of mass-energy equivalence and the principle of energy conservation.

.

**Appendix: Quaternionic Mass Operator and the hypercomplex space**

The operator $P_\mu P^\mu$ is associated with $M^2$ as suggested by the classical relativistic expression $E^2 = p^2 + M^2$. Also, the irreducible representations of the Poincare group are labeled by the eigenvalues (k,s) of the invariant operators and the sign of the energy ±E. Thus if $M^2$ denotes the "square of the mass" then in representations $P^{(k,s)}$ we have $M^2 = \hbar^2 k^2 / c^2$, while in $P^{(0,m)}$ we have $M^2 = 0$ and for space-like **k**, $M^2 < 0$. With this interpretation, the mass would be imaginary for space-like **k**, would be zero for $P^{(0,m)}$ and would take the value $M = \hbar k / c$ for $P^{(k,s)}$ [9].

We can obtain a unified interpretation of the Casimir operator $P_\mu P^\mu = E^2 - \mathbf{p}^2$ (in association with the mass M), which includes the above-named possibilities as special cases. We write

$$P_\mu P^\mu = M^2 = (\beta E)^2 + (\mathbf{\alpha}\cdot\mathbf{p})^2 \; ; (\alpha_i)^2 = -1, \; \beta^2 = 1 \qquad (A.1)$$

where, in the above, the **α**'s are like the Dirac matrices $\mathbf{\alpha}_D$ which are 4 x 4 matrix representations of quaternions. The right hand side of eqn (A.1) can also be put in the form

$$M^2 = \{\beta E + i_\sigma (\alpha \cdot p)\}\{\beta E - i_\sigma (\alpha \cdot p)\} \; ; (i_\sigma)^2 = -1 \qquad (A.2)$$

If, as described above $i_\sigma$ is analogous to the imaginary unit $i = \sqrt{-1}$, then we can write

---

[1] A complex rest-energy term which describes (in time) the growth or decay of states is physically admissible, and is analogous to the complex relative permittivity / complex optical refractive index, used for describing the absorption of energy in dielectric materials / the absorption of light in conductors.

$$Q = \beta E + i_\sigma (\alpha \cdot p) \ , \ Q^* = \beta E - i_\sigma (\alpha \cdot p) \quad (A.3)$$

then (A.2) may be expressed as

$$M^2 = QQ^* = \|Q\|^2 = \text{constant} \quad (A.4)$$

Therefore, Q from (A.3) may be written as,

$$Q = \|Q\| \exp(i_\sigma \vartheta) \quad (A.5)$$

where,

$$\sin \vartheta = \frac{(\alpha \cdot p)^2}{\|Q\|} = \sqrt{\frac{-p^2}{E^2 - p^2}} \ , \ \tan \vartheta = \sqrt{\frac{-p^2}{E^2}} \ \text{and} \ \|Q\| = \sqrt{E^2 - p^2} = M \quad (A.6)$$

As remarked above, $\alpha_i$ (i = x, y, z) just like the unit quaternions **i**, **j**, **k**, have non-commutative algebra, and in the same way that $(\alpha_i)^2 = -1$, we note that $\mathbf{i}^2 = \mathbf{j}^2 = \mathbf{k}^2 = -1$. The space $\mathcal{H}$ of quaternions (hypercomplex numbers) is isomorphic to that of the complex numbers $\Box$ [10]; hence, a given quaternion variable $q = a + \mathbf{i}b + \mathbf{j}c + \mathbf{k}d$ may be written as,

$$q = a + i_r b_r = \|q\| \exp(i_r \Theta) \quad (A.7)$$

where $q^* = a - i_r b_r$, $i_r = (b\mathbf{i} + c\mathbf{j} + d\mathbf{k})/\|r\|$, $\sin \Theta = \|r\|/\|q\|$, $b_r = \|r\| = \sqrt{b^2 + c^2 + d^2}$, and the norm $\|q\| = \sqrt{qq^*} = \sqrt{a^2 + b^2 + c^2 + d^2}$. It is therefore natural to expect that Q, which also is a quaternion variable, may be expressed in complex form. In equation (A.5), $i_\sigma$ may take the specific form $i_\sigma = (\boldsymbol{\alpha} \cdot \mathbf{p})/|\mathbf{p}|$.

From (A.4), it may be observed that the product $QQ^* = \|Q\|^2 = M^2$ = constant, hence it is a Lorentz invariant. Also in eqn (A.6), $\sin \vartheta$ is either real or imaginary depending on whether $P_\mu P^\mu$ is space-like or time-like. For time-like $P_\mu P^\mu$, $\|Q\| = M$ and $\sin \vartheta$ is imaginary, which implies that $\vartheta = i\chi$, and therefore Q can be real, imaginary or complex depending on the value of $\chi$. For space-like $P_\mu P^\mu$, $\sin \vartheta$ is real, but $\|Q\|$ is pure imaginary and, as a consequence, Q can only lie along the imaginary axes of the hypercomplex space $\mathcal{H}$; furthermore $dE = -|p/E|(h^2 - 1)dp$, which is characteristic of superluminal tachyons having pure imaginary mass $\pm iM$, and for which the momentum p increases as the energy E decreases (note: $\sin\vartheta = h^{-1} \in [-1, 1]$). Thus Q can take values other than the real constants $\pm M$ as the phase $\vartheta$ varies over the hypercomplex space $\mathcal{H}$, while its absolute value $M^2$ remains unchanged. This is reminiscent of the behaviour of the complex variable $z = r \exp(i\varphi)$, the magnitude of which, written as $|z| = (zz^*)^{1/2} = r$ is constant, while its phase $\varphi$ traverses the complex plane.

For the Poincare group, the Casimir invariant operator $P_\mu P^\mu$ is associated with the square of the mass $M^2$ as already noted [9]; it is $M^2$ which is significant for the Poincare group rather than M which is used only to label the irreducible representations. Nevertheless, by expressing $P_\mu P^\mu$ as the product of the quaternion variables Q and $Q^*$, it has been shown that the real constant values ±M are special case of $Q = \|Q\| \exp(i_\sigma \vartheta)$ when $\vartheta = 0$. Thus the mass can, in general, be described by a quaternion Q with a variable phase $\vartheta$ traversing the hypercomplex space $\mathcal{H}$ such that the square of its absolute value $\|Q\|^2 = M^2$, a Casimir invariant of the Poincare group, remains constant while its phase changes; the familiar values of the mass, ±M, ±iM, M = 0 etc, which correspond to specific representations of the Poincare group, are all special cases of the variable Q.

For the operator $Q_S = \beta^* E + \alpha^* \cdot \mathbf{p}$, the mass function derived from the Heisenberg equation of motion, using the Dirac Hamiltonian $H_D$, is of the form [7]

$$Q_S(t) = M[A + B\exp(2i\boldsymbol{\alpha} \cdot \mathbf{p} ct/\hbar)] \qquad (A.8)$$

where A = 1, B = −1, and the phase $\vartheta$ depends on $\boldsymbol{\alpha} \cdot \mathbf{p}$ and the coordinate time t. This corresponds to the relative displacement from the origin of the quantity Q by +M on the real axis, while the phase of the quaternion changes, keeping the length $|Q_B| = M^2$ constant. Alternatively, $Q_S$ in (A.8) can be described as the sum of two quaternions $Q_1$ and $Q_2$, where $Q_1 = |Q_1|$ (a quaternion with vanishing phase $\vartheta = 0$) and $Q_2 = |Q_2|\exp(i\vartheta)$. Analogous to complex variables, $|Q_1| - |Q_2| \leq |Q_1 + Q_2| \leq |Q_1| + |Q_2|$, which for (A.4) corresponds to the extremal, real values 0 and 2M, both of which have corresponding representations of the Poincare group, with the latter describing a composite system

.